\documentclass[num-refs]{wiley-article}

\usepackage{siunitx}
\usepackage{natbib}
\usepackage{amsmath}
\usepackage{amsfonts}
\usepackage{bbm}
\usepackage{graphicx,psfrag,epsf}
\usepackage{enumerate}
\usepackage{adjustbox}

\papertype{Original Article}
\paperfield{Journal Section}

\title{Reinforced Borrowing Framework: Leveraging Auxiliary Data for Individualized Inference}

\author[1\authfn{1}]{Ziyu Ji}
\author[1\authfn{1}]{Julian Wolfson PhD}

\affil[1]{Division of Biostatistics, School of Public Health, University of Minnesota, Minneapolis, Minnesota, 55455, U.S.A}

\corraddress{Ziyu Ji, 420 Delaware St.SE, Minneapolis, Minnesota 55455, U.S.A.}
\corremail{jixxx311@umn.edu}

\presentadd[\authfn{2}]{Division of Biostatistics, School of Public Health, University of Minnesota, Minneapolis, Minnesota, 55455, U.S.A}

\fundinginfo{N/A}

\runningauthor{Ziyu Ji et al.}

\begin{document}

\begin{frontmatter}
\maketitle

\begin{abstract}
Increasingly during the past decade, researchers have sought to leverage auxiliary data for enhancing individualized inference. Many existing methods, such as multisource exchangeability models (MEM), have been developed to borrow information from multiple supplemental sources to support parameter inference in a primary source. MEM and its alternatives decide how much information to borrow based on the exchangeability of the primary and supplemental sources, where exchangeability is defined as equality of the target parameter. Other information that may also help determine the exchangeability of sources is ignored. In this article, we propose a generalized Reinforced Borrowing Framework (RBF) leveraging auxiliary data for enhancing individualized inference using a distance-embedded prior which utilizes data not only about the target parameter, but also uses different types of auxiliary information sources to "reinforce" inference on the target parameter. RBF improves inference with minimal additional computational burden. We demonstrate the application of RBF to a study investigating the impact of the COVID-19 pandemic on individual activity and transportation behaviors, where RBF achieves 20-40\% lower MSE compared with existing methods.

\end{abstract}
\end{frontmatter}

\section{Introduction}
\label{sec:intro}
Due to the increasing availability of high-resolution individual-level data, inference on individual units is of interest to researchers across many disciplines. For example, in the context of mobile health (mHealth), it is feasible to collect substantial amounts of data about individual study participants via continuous sensor-based monitoring over time \citep{steinhubl2015emerging}. However, this intensive monitoring can be burdensome to participants, and hence it may not be possible to collect enough data about each individual to make precise inferences. Researchers have proposed various methods to borrow information from other similar individuals or data sources to increase the precision of individual-level inference, for example by using fusion learning via maximum likelihood \citep{cai2020individualized, shen2020fusion}, cooperative learning based on model linkage graph \citep{zhou2021model}, or leveraging auxiliary summary statistics to improve the inference of an internal study under meta-analysis settings \citep{zhang2020generalized}. 

Recently, multisource exchangeability models (MEM, proposed by Kaizer et al. \cite{kaizer2018multi}) have gained prominence as a data borrowing method, particularly in the context of clinical trials. The popularity of MEM is likely inherited from its close relationship to Bayesian model averaging (BMA) \citep{hoeting1999bayesian,fragoso2018bayesian} as BMA  has been widely used in many different disciplines since first proposed in the 1990s \citep{berry2009bayesian, wright2009forecasting, raftery2005using}. BMA takes the weighted average across multiple Bayesian models in order to achieve comprehensive and robust posterior inference of a target parameter. MEM combines the idea of BMA and the exchangeability-nonexchangeability (EX-NEX) model \citep{neuenschwander2016robust}, and has been applied to improve treatment effect estimation in pivotal or basket clinical trials \citep{hobbs2018bayesian, kaizer2019basket}. The MEM framework has been extended in several directions: Brown et al. \cite{brown2021iterated} proposed iterated MEM (iMEM) which reduced the computational complexity of MEM, allowing it to be applied with a larger number of supplemental sources; Kotalik et al. \cite{kotalik2021dynamic} combined the idea of MEM with regression models and considered treatment effect heterogeneity; Ling et al. \citep{ling2022calibrated} applied capping priors on MEM, which controlled the extent of borrowing by placing a cap on the effective supplemental sample size; and Ji et al. \citep{ji2022flexible} developed data-driven MEM (dMEM) that improved performance by filtering out highly nonexchangeable supplemental sources and incorporating data from a large number of sources in the final inference without substantially increasing computational burden. 

Existing MEM techniques for making posterior inferences about a parameter in a primary source use information about that same parameter in secondary sources to determine how much to borrow from them. For example, if the target parameter is the mean, then the sample means of the same parameter from the secondary sources (and their precisions, calculated using standard deviations) are used to determine the amount of borrowing. However, by only leveraging information on the target parameter, MEM and its current extensions do not consider other data in the primary and supplemental sources that may also be useful in determining how much information to borrow from that source. When the target parameter does not provide accurate and precise information indicating the exchangeabilities of the sources, we could utilize this auxiliary information to improve inference.

In our motivating example, we consider data from the COVID Travel Impact (CTI) Study which investigated the impact of the recent COVID-19 pandemic on individual activity and transportation behaviors. One of the measures of interest is the perceived risk of COVID-19 infection during daily activities estimated by self-reported measurements such as the number of close contacts. Suppose we are interested in making inferences about Person A's mean perceived COVID-19 risk. We also observe two other study participants, Person B and Person C, with limited or rough data for us to confidently determine their ground-truth exchangeability with Person A on the COVID-19 risk. However, Person B and Person C may have other characteristics that make them more or less similar to Person A; for example, perhaps Person B (like Person A) generally works from home while Person C is an essential worker who usually works in-person in high risk areas. In this case, it is more likely that the sample mean risk from Person C is actually closer to that of Person A, despite the vague indication from the target parameter. In this example, the individual-level characteristic "work site" can help us decide how much to borrow from each person as it indirectly contributes information to the inference on infection risk. Further, we may also want to use the information on other measures captured simultaneously with the measure on which we are making inferences. For example, in addition to the self-perceived risk of infection, CTI participants were also asked about the perceived level of congestion (i.e., how "busy" an area was) during daily trips and activities. This subjective measurement is highly correlated with the measurements of infection risk and may better represent the contact level during the activity, so it could influence how we want to adjust the borrowing behavior. 

In this article, we propose a reinforced borrowing framework (RBF) using a distance-embedded prior within MEM which utilizes data not only about the target parameter, but also uses other auxiliary information sources to "reinforce" (i.e., improve) inference on the target parameter. The RBF provides a flexible approach to incorporate different types of auxiliary information into the data borrowing process based on MEM. The "reinforcement" provided by the RBF can yield substantial improvements for individual inference ($~$20\% reduction in MSE for our motivating CTI example). The method is straightforward to implement, poses minimal additional computational burden, and is compatible with existing extensions of MEM such as iMEM and dMEM.

The remainder of the article is structured as follows. Section~\ref{sec:overview} gives a brief overview of BMA and a more detailed presentation of MEM, while introducing the notation used in the rest of the sections. Section~\ref{sec:method} introduces our proposed method using distance-embedded priors and Sections~\ref{sec:method_chara} and~\ref{sec:method_para} demonstrate the prior construction with different types of auxiliary data. Section~\ref{sec:method_inMEM} discusses how to incorporate the reinforced borrowing prior into MEM. Section~\ref{sec:sim} provides a series of simulation studies analyzing the performance of our method under different data environments. Section~\ref{sec:application} illustrates a real-world application of the method on the COVID Travel Impact (CTI) Study. Finally, the article is concluded by a brief discussion in Section~\ref{sec:discussion}.

\section{Overview and Notation}
\label{sec:overview}
We begin with an overview of BMA and MEM, which our proposed method builds on. Just as its name implies, BMA takes the weighted average of multiple Bayesian models in order to get the final inference. Given a single Bayesian model $\Omega_k$ estimating a parameter $\theta$ with observed data $D$, the conditional posterior is $P(\theta \mid D, \Omega_k) \sim P(\theta \mid \Omega_k)P(D \mid \theta, \Omega_k)$. Then, BMA is the weighted average of the $K$ Bayesian models: $P(\theta \mid D)=\sum^K_{k=1}w_k P(\theta \mid D, \Omega_k),$ where the weight is
\begin{equation}
w_k=P(\Omega_k \mid D)=\frac{P(D \mid \Omega_k)P(\Omega_k)}{\sum^K_{i=1}P(D \mid \Omega_i)P(\Omega_i)},
\end{equation}
which is the probability that the model $\Omega_k$ is true given the data. For each $k$, $w_k$ could be calculated by using the marginal likelihood of data $P(D \mid \Omega_k)$, and a prior probability $P(\Omega_k)$ that the model $\Omega_k$ is true. Therefore, under the framework of BMA, the final inference is not only determined by a single model, but jointly considers multiple models, thereby providing the potential to integrate different data sources.

The idea of MEM is directly derived from BMA. Suppose the data $D$ includes one primary source $S_p$ and $H$ supplemental sources $S_1,...,S_H$, and the goal is to make inference on a parameter $\theta(p)$ of the primary data source $S_p$ by borrowing information from the same parameters $\theta(1),\dotsc,\theta(h)$ of the supplemental sources. A secondary source $h$ is said to be \textit{exchangeable} with respect to $\theta(p)$ if $\theta(h) = \theta(p)$; we write $s_h=1$ if source $h$ is exchangeable and $s_h=0$ otherwise. The goal of MEM is to identify the supplemental sources most likely to be exchangeable with the primary source $p$ and borrow most strongly from them in making inferences on $\theta(p)$. In MEM, each Bayesian model $\Omega_k$ is defined by a set of supplemental sources assumed exchangeable with the primary source, which could be defined as a set of values of $s_h$ such as ${s_{k1}=1, s_{k2}=0, ...,s_{kH}=1}$, etc. The prior probability $P(\Omega_k)$ when calculating the posterior weight of model $\Omega_k$ can be written as $P(\Omega_k)=P(S_1=s_{k1}) \times ... \times P(S_H=s_{kH})$. Also, MEM requires $P(D \mid \Omega_k)$ in the calculation of posterior weight, which is expressed by $P(D \mid \Omega_k) = \int P(D \mid \theta(p),\Omega_k)P(\theta(p) \mid \Omega_k)d\theta(p)$. 

When the data follow Gaussian, Poisson, or Binomial distributions with known parameters, there are closed-form marginal likelihood and posterior expressions for the key components of MEM. For example, with a flat prior on $\theta(p)$ and Gaussian data, the posterior distribution of parameter $\theta(p)$ given data $D$ with known variances is: 
\begin{equation}
P(\theta(p) \mid D) = \sum^K_{k=1} w_k P(\theta(p) \mid \Omega_k, D) 
= \sum^K_{k=1} w_k \mathcal{N}\Bigg(\frac{\frac{n_p}{\sigma_p^2}\bar{y}_p+\sum^H_{h=1}\frac{n_h}{\sigma_h^2}\bar{y}_h}{\frac{n_p}{\sigma_p^2}+\sum^H_{h=1}s_{kh}\frac{n_h}{\sigma_h^2}}, \Big(\frac{n_p}{\sigma_p^2} + \sum^H_{h=1}s_{kh}\frac{n_h}{\sigma_h^2}\Big)^{-1} \Bigg),
\end{equation}
where the sample means $\bar{y}_p$ and $\bar{y}_h$ are the estimators of the target parameter in the primary source $p$ and supplemental source $h$ respectively; $n_p$ and $n_h$ are the numbers of observations; $\sigma_p$ and $\sigma_h$ are the assumed known standard deviations of the target parameter. When calculating the posterior weights, we usually choose equal prior probabilities on all models, so the posterior weights are proportional to the marginal likelihood $P(D \mid \Omega_k)$ and have closed-form expressions (see \cite{kaizer2018bayesian}). Maximum likelihood-based empirical estimates of $\sigma_p$ and $\sigma_h$ can be used; extensions that treat the covariance matrix as unknown have been published \citep{kotalik2022group, kaizer2018multi}.

\section{Reinforced Borrowing Framework: Using a Distance-embedded Prior in MEM}
\label{sec:method}

In this section, we describe our proposed method, the Reinforced Borrowing Framework (RBF), which utilizes information beyond the target parameter to help refine the parameter inference from MEM. The basic idea of the RBF is that the similarity of supplemental sources with the primary source on aspects \textit{other} than the target parameter can be used to modify (i.e., "reinforce") the standard MEM model weights that are based on the apparent exchangeability of the target parameter. As we show, reinforcing the weights in this way can not only improve efficiency by shrinking the posterior variance but can also correct potential selection bias in the primary source. 

We distinguish between two types of auxiliary information that can contribute to our borrowing procedure:  information about \textit{auxiliary characteristics} and information about \textit{auxiliary parameters}. For each \textit{characteristic}, we only have one observation per unit or individual, while for each \textit{parameter} we have multiple observations that are "aligned" (i.e., collected synchronously) with the target parameter. In what follows, for ease of exposition, we will use abuse terminology slightly and use the term "auxiliary parameter" (sometimes just "parameter") to refer to data that could be used to estimate something other than the target parameter.

\begin{figure}[!p]
\begin{center}
\includegraphics[width=5.5in]{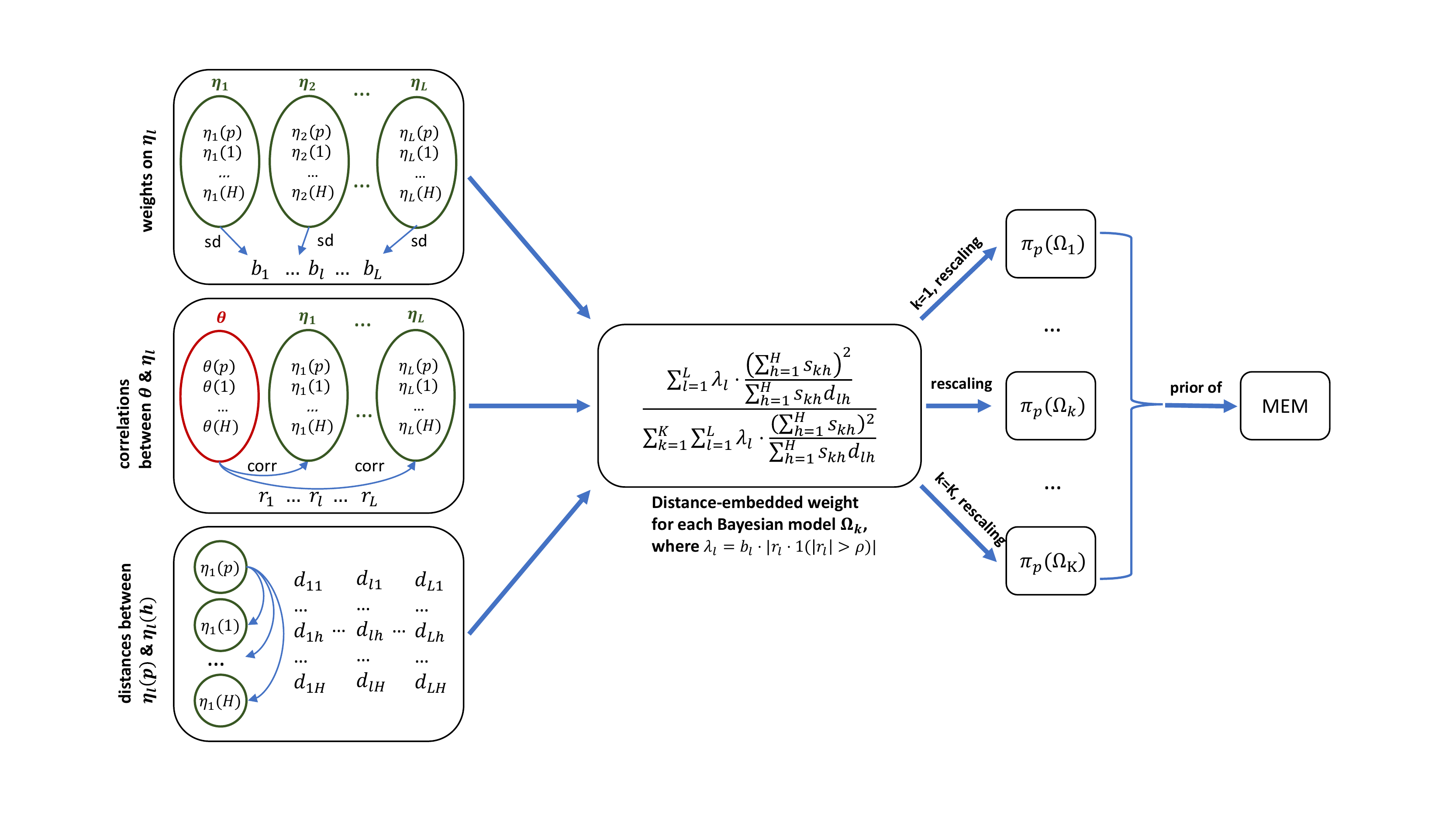}
\caption{Diagram of Distance-embedded Prior for Characteristics in RBF. \label{fig:diagram}}
\end{center}
\end{figure}

A diagram of the method for auxiliary characteristics is shown as Figure~\ref{fig:diagram}, which has notation consistent with expressions in the following subsection. The method for auxiliary parameters is similar to that for characteristics. In our proposed method, we use distance metrics calculated from the auxiliary characteristics or parameters as priors $P(\Omega_k)$ of each Bayesian model, a technique we refer to as a "distance-embedded prior". The main goal of borrowing from those auxiliary data sources is to better determine the exchangeability of the supplemental sources with the primary source when the target parameter provides noisy or misleading information, leading to additional precision, potentially lower bias, and thus better parameter inference. Notice that we want more accurate inferences instead of simply higher precision, meaning that we would also wish to avoid borrowing from the sources that accidentally \textit{appear} to be seemingly exchangeable, although doing so may further increase the estimating precision in an overly confident manner. Sometimes, sources that are determined as exchangeable when only looking at the limited data collected for the target parameter could be not truly exchangeable with the primary source (as the motivating example in Section~\ref{sec:intro} illustrates).

Another advantage of our proposed method is that borrowing auxiliary information could potentially correct selection bias in the primary source affecting the estimation of the target parameter. When the auxiliary characteristics or parameters carry the correct information regarding the true exchangeability of the supplemental sources, inference on the target parameter for the primary source can be driven towards the real underlying distribution by borrowing more from the truly exchangeable supplemental sources. For example, if the target parameter is mean heart rate in beats per minute and the primary source is a person that tends to only test their heart rate when doing physical activity, then the observed distribution of heart rates would be left-skewed. If some characteristics that are less subject to measurement bias (e.g., age, BMI) are highly correlated with heart rate, we can use data on these characteristics to make better decisions about how much to borrow from supplemental sources and thereby correct the skewness in the parameter inference of the primary source. 

\subsection{Distance-embedded Prior for Auxiliary Characteristics}
\label{sec:method_chara}
Auxiliary characteristics are measured at the source level, i.e., they are based on one observation per source. Suppose the goal is to estimate $\theta(p)$ for a primary data source $p$, which also has $L$ observed auxiliary characteristics $\{\eta_1(p),...,\eta_L(p)\}$. Meanwhile, there are $H$ supplemental sources with observations on the target parameter $\theta(1),\dots,\theta(H)$ and the same set of $L$ characteristics $\{\eta_1(1),\dots,\eta_1(H),\eta_2(1),\dots,\eta_L(H)\}$. In each source, multiple outcome samples are observed from the distribution containing the target parameter $\theta$, but each of the characteristics has only one observation per source, e.g. age, income level, BMI, and other one-time qualitative measurements. In a Bayesian model $\Omega_k$, ${s_{k1},...,s_{kH}}\in {0,1}$ are the indicators of source exchangeability, where $s_{kh}=1$ means supplemental source $h$ is assumed to be exchangeable with the primary source on $\theta_p$. Characteristics $\eta_l$ are assumed to be correlated with $\theta$ with a correlation $r_{l}$. 

Specifically, we modify the prior on posterior weights using the distance between the primary and supplemental characteristics. Suppose the distance between primary and supplemental source $h$ on characteristic $\eta_l$ is noted as $d_{lh}$, then the distance-embedded prior for Bayesian model $\Omega_k$ (provided $\sum^H_{h=1}s_{kh} > 0$) is:
\begin{equation}
    \pi_d(\Omega_k)= \frac{\sum^L_{l=1} \lambda_l\cdot\frac{(\sum^H_{h=1}s_{kh})^2}{\sum^H_{h=1}s_{kh} d_{lh}}}{\sum^K_{k=1}\sum^L_{l=1} \lambda_l\cdot\frac{(\sum^H_{h=1}s_{kh})^2}{\sum^H_{h=1}s_{kh} d_{lh}}}, \ 
    \lambda_l=b_l\cdot|r_{l} \cdot \mathbbm{1}(|r_{l}|>\rho)|
    \label{eq:distprior}
\end{equation}
where $b_l$ is the weight for $\eta_l$ to normalize the potentially different scales of the characteristics; $\rho$ is a predetermined positive threshold on the correlation that only keeps auxiliary characteristics that have larger absolute correlations with the target parameter, more details are discussed in section\ref{sec:method_inMEM}.

Clearly, the prior is normalized using the denominator to be added up to 1 for all $k$. The numerator of the equation \eqref{eq:distprior} is represented by taking the summation for the multiplication of two parts: the weighting term $\lambda_l$ and the distance term. The weighting term contains the weights $b_l$ on $\eta_l$ to standardize different characteristics, as well as the correlation between $\theta$ and $\eta_l$ to quantify the relationship between the parameter of interest and characteristics. The distance part is derived from the number of assumed exchangeable sources for the corresponding Bayesian model times the inverse of the average distance between the primary source and the exchangeable sources, which symbolizes the similarity between the primary source and the selected (meaning that $s_{kh}=1$) supplemental sources of the Bayesian model $\Omega_k$ in a normalized way.

The prior in \eqref{eq:distprior} is not well-defined when the Bayesian model $\Omega_k$ has $\sum^H_{h=1}s_{kh} = 0$ thus $s_{k1},...,s_{kH}=0$, which means none of the supplemental sources are assumed exchangeable with the primary source. So, we apply a flat prior to this circumstance to complete the prior specification, which means the prior is represented as $\frac{1}{2^H}$ when $\sum^H_{h=1}s_{kh} = 0$. The finalized distance-embedded prior for any $\Omega_k$ is: 
\begin{equation}
\pi_d(\Omega_k)= \mathbbm{1}\left(\sum^H_{h=1}s_{kh} > 0\right)\cdot\frac{2^H-1}{2^H}\cdot\frac{\sum^L_{l=1} \lambda_l\cdot\frac{(\sum^H_{h=1}s_{kh})^2}{\sum^H_{h=1}s_{kh} d_{lh}}}{\sum^K_{k=1}\sum^L_{l=1} \lambda_l\cdot\frac{(\sum^H_{h=1}s_{kh})^2}{\sum^H_{h=1}s_{kh} d_{lh}}} + \mathbbm{1}\left(\sum^H_{h=1}s_{kh} = 0\right)\cdot\frac{1}{2^H}.
\end{equation}

Parameters $d,\lambda, b, r$ are estimated and plugged into the formula in a data-driven manner. Usually, we use squared euclidean distance (SED) as the distance metric: $\hat{d}_{lh} = (\hat{\eta_l}(h) - \hat{\eta_l}(p))^2$, where $\hat{\eta_l}(h)$ and $\hat{\eta_l}(p)$ are the observed characteristics $\eta_l$ for the primary source and supplemental source $h$, respectively. In one of our primary studies, when comparing with euclidean distance, adding up the squared euclidean distance of assumed exchangeable sources had better performance in matching the ground truth exchangeability status of the Bayesian models. The framework is also applicable to other distance metrics if they are better fits. The weight for characteristics could be constructed using either the pooled standard deviation ratio: $\hat{b}_l = \hat{\sigma_l}/\sum^L_{k=1}\hat{\sigma_k}$, where $\hat{\sigma_k}$ is the pooled standard deviation of characteristic $\eta_k$ with observations from all the sources (including the primary source); or the inverse of pooled variance $\hat{b}_l = 1/\hat{\sigma_l}^2$. When using the inverse of pooled variance as the weight for characteristics, the constructed distance between sources turns out to be the squared Mahalanobis distance. The performance of the two weights is similar, so we will use the pooled standard deviation ratio as a default in this paper. In addition, if the target parameter is mean, we use Pearson's correlation coefficient as the correlation estimator $\hat{r}_{l}$, while RBF is also flexible to other correlation estimators or even other ways to quantify the similarity between two variables. If selecting an alternative estimator other than mean, we suggest considering the correlation estimator between the target parameter and auxiliary characteristic correspondingly. For example, if the estimator is minimum or maximum, it's better to choose Spearman's correlation over Pearson's correlation since Spearman's correlation is based on the ranked values rather than the raw data, thus more consistent with estimators related to ranking.

\subsection{Distance-embedded Prior for Auxiliary Parameters}
\label{sec:method_para}
We assume that there are multiple observations contributing to only estimating the target parameter in either the primary or supplemental sources. In some cases, there may also be additional observations that could contribute to estimating other parameters that are potentially correlated with the target parameter. We refer to these as \textit{auxiliary parameters}. Note that the distinction between auxiliary characteristics and auxiliary parameters is the number of per-source observations used to define/estimate them; an auxiliary characteristic is defined by a single observed value per source, while an auxiliary parameter is estimated based on multiple observations per source. We could borrow information from $L$ auxiliary parameters, noted as ${\zeta_1,...,\zeta_L}$. Parameter $\zeta_l$ is assumed to have correlation $r_{l}$ with $\theta$ and the distance between primary and supplemental source $h$ on parameter $\zeta_l$ is noted as $d_{lh}$. For each source, we have vectors of measures on the auxiliary parameters, and there are two ways to define the distance function. First, the distance could be defined by distance metrics between distributions, such as Kullback-Leibler (KL) divergence, Hellinger distance, or Bhattacharyya distance, which all belong to the family of f-divergence. Among those distance metrics, KL divergence is the most commonly used, and a symmetric version of KL divergence is Jeffreys divergence, which is used in our distance function: 
\begin{equation}
\hat{d}_{lh} = D_{KL}(\hat{\zeta}_l(p)||\hat{\zeta}_l(h))+D_{KL}(\hat{\zeta}_l(h)||\hat{\zeta}_l(p)),
\end{equation}
where $\hat{\zeta}_l(p)$ and $\hat{\zeta}_l(h)$ are the observed probability density function of the parameter $\zeta_l$ on primary source and supplemental source $h$, respectively. Hellinger distance and Bhattacharyya distance are also both symmetric and could aslo be potentially applied as the distance function. On the other hand, the parameters could be also treated as characteristics by collapsing to the form of a single observation per source. When collapsing the original auxiliary parameter, as the default, we would recommend using the same estimator as the target parameter, or the users should prudently use their best judgment and expertise on the specific practical circumstance to select appropriate estimators that represent the exchangeability in the target parameter. The corresponding estimated correlations between the parameters and distances between the primary source and supplemental sources could be then calculated in a similar manner to the characteristics described in Section 3.1.  


\subsection{Distance-embedded Prior in MEM}
\label{sec:method_inMEM}
When applying the prior on MEM, we seek to balance between the amount of borrowing from the target parameter and the amount of borrowing from auxiliary characteristics or parameters. Hence, we propose a new prior for model $\Omega_k$ which is a convex combination of the original flat prior and the constructed distance-based prior:
\begin{equation}
    \pi(\Omega_k)= (1-a)\cdot\frac{1}{2^H}+a\cdot\pi_d(\Omega_k).
    \label{eq:demem}
\end{equation}
$\pi(\Omega_k)$ can be plugged into the formula that calculates $w_k$ for MEM. 

The value of $a \in [0,1]$ in \eqref{eq:demem} determines the relative influence of the original and  distance-based priors. Usually, we recommend selecting $a=1$ to fully leverage the benefit of information borrowing. However, when the correlation between the target parameter and auxiliary characteristics or parameters $r_l$ is small, the distance-embedded prior will introduce bias into the model and influence the performance of the final MEM. There are two solutions for those cases: we could either select $a<1$ to borrow less from the auxiliary information and control the loss caused by the extra bias, or set $\rho>0$ to drop the parameters which are weakly correlated with the target parameter and have a high risk of introducing unnecessary bias into the model. In order to maximize the gain and minimize the potential loss, we recommend applying the second approach. A rule-of-thumb selection of $\rho$ based on our simulations is 0.3, as little seems to be gained when auxiliary characteristics or parameters have lower correlations than this. If there is particular concern about weakly correlated characteristics potentially doing more harm to the final performance when there is selection bias in the primary source, we recommend increasing $\rho$ accordingly as an adjustment. 

In addition, we apply normalization on the auxiliary characteristics (similarly for parameters) with pooled min and max before information-borrowing to make sure the characteristics are on the same 0-1 scale. Note that the usual standardization is not appropriate in our case because we assume that our data arise from a mixture of normals. Thus, when there are both exchangeable and nonexchangeable sources, conventional location-scale normalization generally assumes sometimes over-shrinks the data, which may lead to the distance calculation also being nonlinearly shrunk undesirably for some of the characteristics or parameters. 

\section{Simulation Studies}
\label{sec:sim}
The goal of our simulation studies is to demonstrate the performance of our proposed distance-embedded prior. Via three different simulation scenarios, we show in this section that our proposed approach not only achieves better performance over the regular MEM with flat prior on posterior weights when using auxiliary characteristics or auxiliary parameters, but can also correct selection bias in the primary source through the data borrowing process. The reinforced borrowing framework provides the most benefit for the inference of a target parameter when there is a limited amount of primary data available which can be "reinforced" with auxiliary characteristics and parameters. So, our simulation settings are designed to reflect such scenarios. To evaluate the methods, the measures include posterior variances, biases, mean square errors (MSE) or root mean square errors (RMSE), and the posterior weight of the ground-truth 'correct' model. We note that our approach is not expected to yield a large increase in the effective supplemental sample sizes \citep{morita2008determining} (ESSS) over regular MEM because we are mainly directing the method to borrow from the appropriate sources instead of borrowing more. Results on ESSS are available as supplemental Materials. 

For all scenarios, the simulations are repeated 1000 times with different random seeds in data generation. To better demonstrate the basic behavior of our method, we set $\rho=0$ between the target parameter and auxiliary characteristics or parameters (as we recommend in Section \ref{sec:method_inMEM}) in Simulation I and II. In Simulation III, we do not change the correlations and provide two scenarios with $\rho=0.5$ as comparisons. All calculations are completed with R version 4.0.2 \citep{R}.

\subsection{Simulation Scenario I: Borrowing from Characteristics}
\label{sim1}

The first scenario illustrates the case when the distance-embedded prior is constructed based on auxiliary characteristics. The characteristics are assumed to be correlated with the target parameter but not necessarily with each other because it is common to take the characteristics of the sources as observed. In addition, although the covariance matrix among the characteristics is not strictly a diagonal matrix and the correlation between characteristics would cause duplicated information if characteristics are considered independent, the correlation would affect all supplemental sources equally, which just multiplies a scale on all the distance metrics and would have minimal impact on the final weight. The target parameter is from either $N(0,1)$ for the primary source and 5 exchangeable supplemental sources or from $N(1,1)$ for 5 nonexchangeable supplemental sources. The sample size of the primary source is fixed at 10, while supplemental sources have sample sizes from 5 to 15. The auxiliary characteristics are generated independently with predetermined correlations $r_l$ with the ground truth of the source-level means of the target parameter. More details about generating the auxiliary characteristics are available in the supplemental Material.


\begin{figure}[!p]
\begin{center}
\includegraphics[width=5.5in]{sim1.pdf}
\caption{Simulation I Results: Comparing the Performance of RBF versus MEM. Borrowing information from Characteristics using Distance-embedded Prior. Measuring metrics including posterior variance, bias, RMSE and posterior weight of the correct model. The horizontal short lines stand for the median and the vertical bars are extended from 25th percentile to 75th percentile. The names of each bar are the correlations used in generating the characteristics. Scenarios with higher comparability are placed within the same slot separated by dashed lines.  \label{fig:sim1}.}
\end{center}
\end{figure}

Different correlation combinations between the target parameter and characteristics are tested in this scenario. The results are shown in Figure~\ref{fig:sim1}. Each bar in the figure stands for a separate scenario with different sets of correlations between the characteristics and the means of the target parameter. Directly from the plot, almost all the scenarios have better performance than the original MEM, except the case with characteristics all weakly correlated with the target parameter. In the best case with correlations of 0.99, 0.7 and 0.5, the distance-embedded prior decreases posterior variance by a median of 36.8\%, decreases bias by 55.5\%, leading to a reduction in MSE of 41.4\% (RMSE lower by 23.5\%). The posterior weight of the correct model increases by 68.8\%. When comparing across different scenarios, the performances are close between having 3 characteristics with a correlation of 0.7 and having 5 characteristics with the same correlations. In general, the higher the correlation between the auxiliary characteristic and the target parameter, the more advantage could be obtained by the distance-embedded prior. Weakly correlated characteristics (correlation $\leq$ 0.3) appear to harm the performance in terms of posterior, bias and RMSE, but if there are also characteristics with high correlation in the study, the effect would be offset to some extent. For example, in one scenario the advantage gained from a characteristic with a correlation of 0.7 compensate for the disadvantage introduced by two characteristics with correlations 0.3 and 0.1, with 9.5\% lower posterior variance, 7.9\% lower bias, and have MSE around 9\% lower. Also, even when the characteristics are only weakly correlated with the target parameter, the weight of the correlated model is still increased by 43.4\%, which means the prior keeps directing the MEM towards the ground truth. 

\subsection{Simulation Scenario II: Borrowing from Auxiliary Parameters}
\label{sim2}

Under this scenario, we test the performance of the distance-embedded prior when borrowing from auxiliary parameters. For simplicity, we assume that there are the same sample sizes from each of the auxiliary parameters, and that this number matches the sample size from the target parameter. We generate the target parameter of a primary source and 5 exchangeable supplemental sources from $N(0,p(1-p))$ and generate the 5 nonexchangeable sources from $N(1,p(1-p))$. We randomly draw 10 samples for the primary source and the sample sizes of supplemental sources are randomly selected from 5 to 15. Then $p$ is dynamically calculated from the sample sizes as the proportion of the exchangeable samples to make sure the correlations are correctly preserved when generating auxiliary parameters using the Cholesky decomposition. Details are described in the Supplemental Material. The distance function used in this study is Jeffery's divergence since it provides the most consistent results compared with other divergences mentioned before. 

\begin{figure}[!p]
\begin{center}
\includegraphics[width=5.5in]{sim2.pdf}
\caption{Simulation II Results: Comparing the Performance of RBF versus MEM. Borrowing information from Parameters using Distance-embedded Prior. Measuring metrics including posterior variance, bias, RMSE and posterior weight of the correct model.The horizontal short lines stand for the median and the vertical bars are extended from 25th percentile to 75th percentile. The names of each bar are the correlations between the target parameter and the parameters, details about other elements in the correlation matrix is available as supplemental Material. Scenarios with higher comparability are placed within the same slot separated by dashed lines. \label{fig:sim2}}
\end{center}
\end{figure}

The results are illustrated in Figure~\ref{fig:sim2}. Note that the numbers in the title of each bar are the correlations between the target parameter and the parameters; the full covariance matrices are in Table 1 in the supplemental Material. The performance of auxiliary parameters has a similar trend as for auxiliary characteristics, although the magnitudes are smaller. The best performance scenario is still the case with correlations 0.99, 0.7, and 0.5, which has 31.9\% lower posterior variance, 19.1\% lower bias, and 20.4\% lower MSE (10.8\% lower RMSE) when compared with a regular MEM. More parameters with the same correlation of 0.7 provides only modest benefit to the final MSE (12.4\% lower for 3 parameters vs 14.5\% lower for 5 parameters). The shape of the auxiliary parameter distribution also does not have much impact on the result when we generate the parameters from either an exponential distribution or normal distribution. The parameters with lower correlations do not hurt the performance as much as the scenario for characteristics, but there is barely any advantage by borrowing from them. Still, the weights on the correct model are always higher for the distance-embedded prior. 

\subsection{Simulation Scenario III: Correcting the Selection Bias of the Target parameter Using Characteristics}
\label{sim3}

To illustrate the method without making the problem overly complicated, we consider the selection bias of the primary source as represented by the truncated normal distribution. The target parameter of supplemental sources is either from regular or truncated normal distributions, with randomly sampled truncating thresholds. The correlations of the 3 characteristics are set to be $0.99, 0.7$ and $0.5$, respectively. Note that the characteristics are correlated with the means of the ground-truth unbiased target parameter instead of the observed means of the truncated normal distributions. The bias in the results is also calculated with respect to the true underlying mean of 0. 

\begin{figure}[!p]
\begin{center}
\includegraphics[width=5.5in]{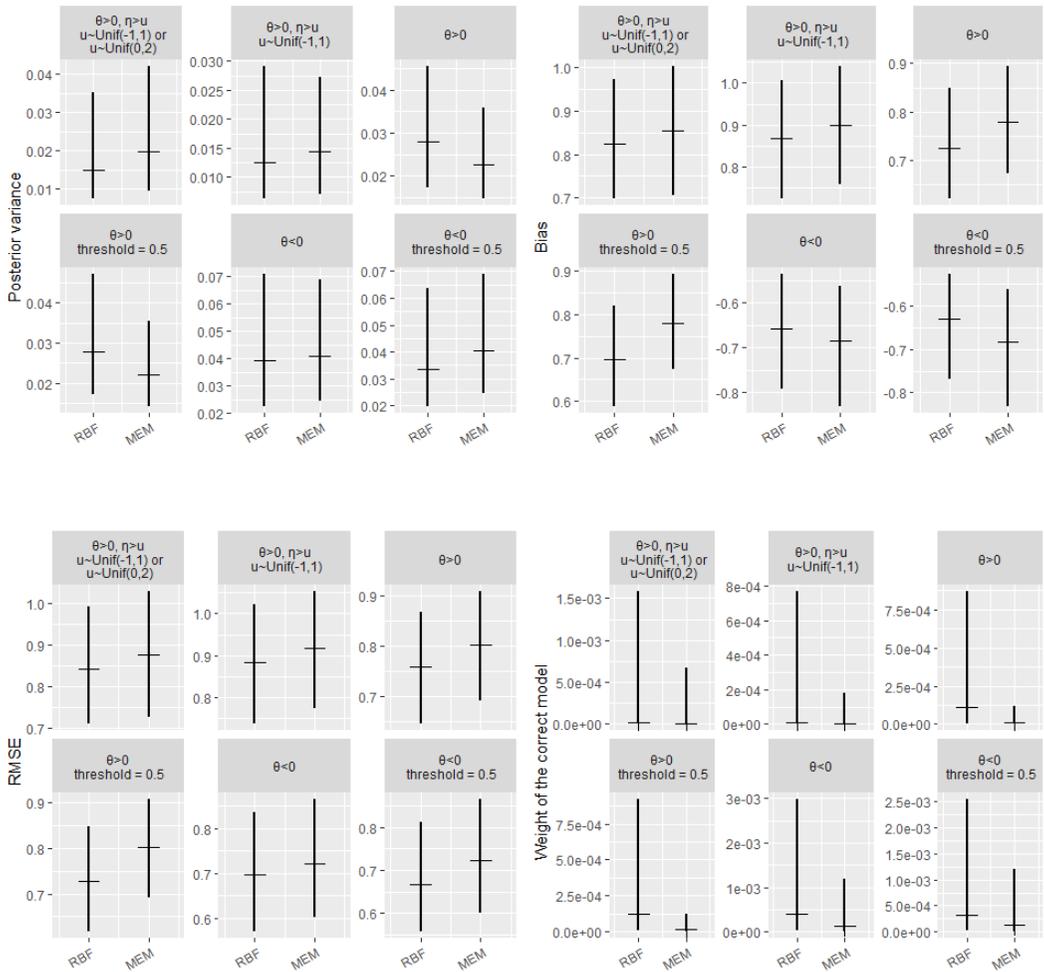}
\caption{Simulation III Results: Comparing the Performance of RBF versus MEM. Borrowing information from characteristics using distance-embedded prior on data with selection bias. Measuring metrics including posterior variance, bias, RMSE and posterior weight of the correct model. The horizontal bars stand for the median and the vertical lines are extended from 25th percentile to 75th percentile. Each panel is a scenario. There are 4 basic scenarios in this simulation study: $\theta>0$ means we have the target parameter of the primary source sampled from truncated normal $TN_{x>0}(0,1)$; $\theta<0$ means the target parameter of the primary source is from $TN_{x<0}(0,1)$; '$\theta>0, \eta<u, u\sim Unif(-1,1)$' is when the target parameter of the primary source from truncated normal $TN_{x>0}(0,1)$, the target parameter of the supplemental sources from truncated normal $TN_{x>u}(0,1)$ if exchangeable otherwise $TN_{x>u}(1,1)$, where $u\sim Unif(-1,1)$; and '$\theta>0, \eta<u, u\sim Unif(-1,1)\ or\ u\sim Unif(0,2)$' is a similar case with the target parameter of the exchangeable supplemental sources have $u\sim Unif(-1,1)$ and non-exchangeable supplemental sources have $u\sim Unif(0,2)$. There are two extra cases with $\rho=0.5$ for $\theta>0$ and $\theta<0$. Note that the y-axes are free-scaled.  \label{fig:sim3}}
\end{center}
\end{figure}

We test different truncating thresholds and directions in different scenarios. The results are shown in Figure~\ref{fig:sim3}. There are 4 basic scenarios in this simulation study: $\theta>0$ means the target parameter of the primary source is sampled from a truncated normal $TN_{x>0}(0,1)$; $\theta<0$ means the target parameter of the primary source is from $TN_{x<0}(0,1)$; '$\theta>0, \eta<u, u\sim Unif(-1,1)$' is when the target parameter of the primary source is from a truncated normal $TN_{x>0}(0,1)$, the target parameter of the supplemental sources from truncated normal $TN_{x>u}(0,1)$ if exchangeable otherwise $TN_{x>u}(1,1)$, where $u\sim Unif(-1,1)$; and '$\theta>0, \eta<u, u\sim Unif(-1,1)\ or\ u\sim Unif(0,2)$' is a similar case with the target parameter of the exchangeable supplemental sources having $u\sim Unif(-1,1)$ and non-exchangeable supplemental sources having $u\sim Unif(0,2)$. There are two extra cases with $\rho=0.5$ for $\theta>0$ and $\theta<0$. Among all the scenarios, $\theta>0$ with $\rho=0.5$ works the best with a 10.8\% decrease in bias and 17.7\% reduction in MSE, compared with a 6.9\% decrease in bias and 10.6\% reduction in MSE for the case with $\rho=0$. The changes in bias and MSE are substantial in absolute values because the bias from the original MEM is around 0.8. The posterior variance of both $\theta>0$ scenarios with or without $\rho=0$ are inflated by over 20\%, but we consider that the increase is actually a sign of good performance, because the original MEM would borrow more than it should from non-exchangeable sources, and lead to an artificially small posterior variance. The posterior weights of the $\theta>0$ cases are 7 to 8 times the original MEM, meaning that the distance-embedded prior greatly encourages borrowing from truly exchangeable sources and discourages borrowing from nonexchangeable sources that appear to be exchangeable. The idea is different for the $\theta<0$ scenario when we try to encourage borrowing in general because none of the sources appears to be non-exchangeable and the null model takes most of the posterior weights. We have posterior variances decreased by 17.4\% and 3.4\% respectively for $\rho=0.5$ or $0$. The decreases in bias are 7.8\% and 4.1\%, and the reductions in MSE are 14.5\% and 6.7\%. For the other two scenarios with biased supplemental sources, the method is robust to the two different ways of truncation for supplemental sources but both of the cases only have modestly around 3.5\% improvements on bias and around 7.3\% deductions on MSE, while there the posterior variances decline by 13.0\% and 25.1\%, respectively.

\section{Application}
\label{sec:application}
During the past two years, daily lives and routines have been reshaped by the COVID-19 pandemic. In order to develop policies that improve the quality of life during and after the pandemic, the COVID Travel Impact (CTI) Study was conducted to investigate the impact of the pandemic on people's daily trips and activities. CTI enrolled 160 participants living in the Minneapolis-St. Paul area that successfully completed the intake survey followed by 14 days of data collection using a mobile application, Daynamica, that records daily trips and activities using smartphone sensor data and delivers surveys asking participants to provide additional COVID-related information about those trips and activities. In an end-of-day survey, participants were prompted to provide information about their trips and activities during the day.  

In our study, to further understand the risk of exposure in different subpopulations, we are interested in how to better estimate the number of contacts during activities. In event surveys collected during the day, there are questions asking participants to report the approximate number of contacts (as an integer) and the level of congestion or crowdedness (on a 1-4 scale) during each trip or activity. In the end-of-day survey, the participants were asked about their level of concern about having contracted COVID that day (on a 1-5 scale). For this illustration, our target parameter is the mean number of contacts during activities for a given individual. All other information is transformed to characteristics by taking the individual-level average, which means for each question, either from the event surveys or the end-of-day survey, there is only one number per person to borrow from. Also, to further highlight the potential value of our method when used in conjunction with MEM, we focus on the subpopulation of $n=11$ participants who only work from home (having $n=10$ supplemental sources for each primary source makes standard MEM computationally feasible, though as we note in the Discussion, RBF is also compatible with approaches that do source selection for MEM). Our goal is to do inference on the mean for each of the 11 participants, so all individuals would serve as the primary source. With a specific primary source, we randomly subsample 10 measures on the target parameter respectively for all sources. The process is applied repeatedly 500 times with the original MEM, which only borrows from the target parameter, and RBF, which borrows from both the characteristics and the target parameter, on each participant to be treated as the primary source. A naive approach that directly calculates the sample mean and the standard deviation is also applied as a comparison. Through leveraging the estimated results from the 500 iterations on 11 participants, the posterior variance, bias, MSE, and ESSS are calculated to compare the performance of the three approaches. Note that bias is estimated as the posterior mean minus the 'ground-truth' mean of all the observations from the individual. There are on average 73 total observations per individual, which is usually enough to achieve a good parameter inference, so we assume that the general mean from all the observed data is the ground truth of the target parameter.

\begin{figure}[!p]
\begin{center}
\includegraphics[width=5in]{app.pdf}
\caption{Application Results: (a) Average sum of posterior weights from models containing the supplemental source (only available for RBF and MEM). For each user, the colors in the figures represent the average sum of posterior weights from the models containing the supplemental source in y-axis across 500 iterations. Each user has 0 weight with itself. (b) Comparing the performance of RBF versus MEM and the naive approach. The target parameter is the average number of contacts during activities. Measuring metrics including posterior variance, bias, RMSE and ESSS. The differences in medians of posterior variance, absolute bias, MSE and ESSS are statistically significant.  \label{fig:app}}
\end{center}
\end{figure}

The summarized results are shown in Figure~\ref{fig:app}. In the top figure, it shows the posterior weights in RBF and MEM. On the x-axis, each column represents the results of borrowing from a single individual viewed as the primary source and the colors represent the average sum of posterior weights from the models containing the supplemental source in the y-axis across 500 iterations. The colors are clearly changed and the RBF figure appears to have higher contrast, meaning that the data borrowing process is reinforced in RBF by dynamically either upweighting or downweighting certain supplemental sources for different borrowers. In the figure below, RBF leads to lower posterior variance, lower bias, lower MSE, and a lightly higher ESSS compared with both the original MEM and the naive approach. RBF has a 27.4\% reduction in the median posterior variance compared with the original MEM (0.093 vs 0.128) and a 3.1\% decrease in the median absolute bias (0.356 vs 0.368), which leads to an 18.0\% decline in the median MSE (0.341 vs 0.416). All differences in performance metrics substantially exceed the magnitude that would be expected by chance given our stochastic resampling process. The median ESSS of the RBF is 10.6\% larger than MEM (26.047 vs 23.553) and as expected, the resulting disparity is even larger between RBF and the naive approach. 

\section{Discussion}
\label{sec:discussion}

In this paper, we have proposed the reinforced borrowing framework (RBF) as a novel comprehensive data leveraging method inspired by multisource exchangeability models (MEM). By combining data not only from the supplemental sources on the same targeted parameter, but also from more alternative data sources, such as external or auxiliary parameters and characteristics which are closely correlated with the targeted parameter, RBF increases the efficiency of individual parameter inference and decreases the potential for selection bias. When compared with the previous application of MEM on heterogeneous treatment effects \citep{kotalik2021dynamic}, RBF avoids joint modeling when leveraging high dimensional data. In addition, RBF is easily compatible with the current extensions of MEM such as iMEM \citep{brown2021iterated} and dMEM \citep{ji2022flexible} for source selection and clustering when there are a large number of available supplemental sources. Overall, RBF is a flexible and efficient data borrowing framework that outperforms the existing methods in a range of scenarios.

In a secondary simulation study (not shown in this manuscript), we found that under certain simulation settings, the approach of collapsing auxiliary parameters to their means and then treating parameters the same way as characteristics may sometimes significantly outperform using distribution-based distance metrics, such as f-divergences. Under our simulation setups, the divergences tend to overreact to the difference in the shape or spread of the distributions between the primary and supplemental sources, especially when the sample size is small and the sample SD estimation is sensitive to the sample selection. For instance, the divergence between the primary and a supplemental source might be overestimated when there are outliers in the observation and the SD estimation is inflated. However, we still believe in the potential advantage of using divergences as distance metrics when the shape or spread of the auxiliary parameter distribution also implies the exchangeability or non-exchangeability of the target parameter. Because when the standard deviation of the auxiliary parameter is large enough to cover the signal of exchangeability in the mean or median, using the sample mean of the auxiliary parameter may cause wrongly estimated correlations and mislead the borrowing process. However, the advantages and disadvantages could be conditional on the specific scenario. In general, we recommend exploring the distribution of both parameters of interest and auxiliary parameters before applying the method.

RBF also has some limitations in its current form. First, RBF is now demonstrated only for estimating the mean. It could be challenging to derive a closed-form posterior distribution for some of the other parameters, and an MCMC process may substantially increase computational complexity. This should not be considered a weakness specifically for RBF since similar limitations apply to almost all Bayesian-based methods. Second, the advantage of RBF over MEM on posterior variance and MSE is more clear when the available sample size for individual sources is modest. When there is more data, parameter inference does not require as much "reinforcement" and hence the extra information borrowed by RBF does not make much difference. However, as we show, even in larger sample size scenarios RBF could still provide value when selection bias impacts the data available for estimating the target parameter. Third, to provide substantial reinforcement of data borrowing, auxiliary characteristics and parameters must be moderately to highly correlated with the target parameter. In practical problems, there will not always be measures available with the high degree of correlation needed to achieve the largest possible gains from RBF.

\section*{Supporting Information}
\subsection*{Data Generation in Simulation}
\subsubsection*{Simulation Scenario I: Borrowing from Characteristics}
The parameter of interest is from either $N(0,1)$ for the primary source and 5 exchangeable supplementary sources or from $N(1,1)$ for 5 nonexchangeable supplementary sources. The sample size of the primary source is fixed at 10, while supplementary sources have sample sizes from 5 to 15. The auxiliary characteristics are generated independently with predetermined correlations $r_l$ with the ground truth of the source-level means of the parameter of interest, which is a vector of 6 zeros and 5 ones, noted as $Y$. For each characteristic $X_l$, we first randomly sample a vector $X_l'$ from a Gaussian distribution, and the correlated characteristic is $X_l=r_l \cdot SD(residual(Y)) \cdot Y + \sqrt{1-r_l^2} \cdot SD(Y) \cdot residual(Y)$, where $residual(Y)$ is a vector that removes the component of $Y$ from the $X_l'$ and becomes orthogonal to $Y$. Note that the generated characteristics would almost always have lower estimated correlations in practice compared with the desired correlations because of the randomness of the parameter of interest.

\subsubsection*{Simulation Scenario II: Borrowing from Auxiliary Parameters}

For simplicity, we assume that there are the same number of observations from each of the auxiliary parameters, and that this number matches the number of observations from the parameter of interest. In order to generate parameters with fixed correlation matrix $R$ and covariance matrix $\Sigma_R$, where $\Sigma_R= Diag(\sigma) \cdot R \cdot Diag(\sigma)$, where $Diag(\sigma)$ is a diagonal matrix with the desired standard deviation of parameters as diagonal elements, a common way is using Cholesky decomposition of $\Sigma_R=C^TC$. Suppose all the observations of our parameter of interest could form a random column vector $Y$, to get the matrix with auxiliary parameters as columns $X$ such that the full parameter matrix $(Y, X)$ has correlation matrix $R$, we could first generate each row of random samples $X'$ identically and independently with the same size as $X$. Then, the parameter matrix could be obtained by $(Y,X) = (Y,X')C$ when the first element of $Diag(C)$ equal to 1 to preserve the first column and make sure the covariance matrix of $(Y,X)$ will then become $C^TCov((Y,X'))C$, where we could set $Cov((Y,X'))=I$ since the standard deviation of $Y$ is determined to be 1 if the first element of $Diag(C)$ is 1 and the columns of $X'$ are any independent random samples. A positive definite covariance matrix example is as below.
$$\Sigma_R = I \begin{pmatrix}
1 & 0.7 & 0.3 & 0.1\\
0.7 & 1 & 0.4 & 0.1\\
0.3 & 0.4 & 1 & 0.05\\
0.1 & 0.1 & 0.05 & 1\\
\end{pmatrix} I$$

Note that the standard deviation of $Y$ is the pooled standard deviation across all observations from different sources. Thus, we have to adjust the standard deviations proportionally when generating the sources considering the heterogeneity between the sources introduces extra variability to $Y$. Since our parameter of interest is supposed to be Gaussian, the standard deviation of the mixed Normal is $\sigma^2+P(1-p)(\mu_0-\mu_1)^2$, with $p$ as the proportion of the observations to be exchangeable, exchangeable sources from $N(\mu_0,\sigma^2)$ and non-exchangeable sources from $N(\mu_1,\sigma^2)$. So, when given exchangeable mean $\mu_0=0$ and non-exchangeable mean $\mu_1=1$, with a desired standard deviation of the mixed Normal as well as the standard deviation of $Y$ to be 1, we could solve that $\sigma=\sqrt{P(1-p)}$. In practice, we use the actual selected sample sizes to calculate $p$ to keep the theoretical standard deviation of $Y$.

We generate the parameter of interest of a primary source and 5 exchangeable supplementary sources from $N(0,p(1-p))$ and generate the 5 nonexchangeable sources from $N(1,p(1-p))$. There are 10 samples in the primary source and the sample sizes of supplementary sources are randomly selected from 5 to 15. Then $p$ is dynamically calculated from the sample sizes as the proportion of the exchangeable samples. The auxiliary parameters are then generated as described above using Cholesky decomposition. To keep the process simple, we randomly generate each column in $X'$ from $N(0,1)$ or $Exp(1)$, depending on our assumptions on the distribution of auxiliary parameters. The distance function used in this study is Jeffery's divergence since it provides the most consistent results compared with other divergences mentioned before.

\subsubsection*{Simulation Scenario III: Borrowing from Auxiliary Parameters}

The original distribution for the primary source and exchangeable supplementary sources is always $N(0,1)$ and for the non-exchangeable sources is $N(1,1)$. The sample sizes and correlated characteristics are generated similarly as in Simulation Scenario I. 

\printendnotes


\bibliography{ref}

\begin{thebibliography}{22}
\providecommand{\natexlab}[1]{#1}
\providecommand{\url}[1]{\texttt{#1}}
\providecommand{\urlprefix}{}

\bibitem[{Steinhubl et~al.(2015)Steinhubl, Steven R and Muse, Evan D and Topol,
  Eric J}]{steinhubl2015emerging}
Steinhubl SR, Muse ED, Topol EJ.
\newblock The emerging field of mobile health.
\newblock Science translational medicine 2015;7(283):283.

\bibitem[{Cai et~al.(2020)Cai, Chencheng and Chen, Rong and Xie,
  Min-ge}]{cai2020individualized}
Cai C, Chen R, Xie Mg.
\newblock Individualized inference through fusion learning.
\newblock Wiley Interdisciplinary Reviews: Computational Statistics
  2020;12(5):1498.

\bibitem[{Shen et~al.(2020)Shen, Jieli and Liu, Regina Y and Xie,
  Min-ge}]{shen2020fusion}
Shen J, Liu RY, Xie Mg.
\newblock iFusion: Individualized fusion learning.
\newblock Journal of the American Statistical Association
  2020;115(531):1251--1267.

\bibitem[{Zhou et~al.(2021)Zhou, Jiaying and Ding, Jie and Tan, Kean Ming and
  Tarokh, Vahid}]{zhou2021model}
Zhou J, Ding J, Tan KM, Tarokh V.
\newblock Model Linkage Selection for Cooperative Learning.
\newblock J Mach Learn Res 2021;22:256--1.

\bibitem[{Zhang et~al.(2020)Zhang, Han and Deng, Lu and Schiffman, Mark and
  Qin, Jing and Yu, Kai}]{zhang2020generalized}
Zhang H, Deng L, Schiffman M, Qin J, Yu K.
\newblock Generalized integration model for improved statistical inference by
  leveraging external summary data.
\newblock Biometrika 2020;107(3):689--703.

\bibitem[{Kaizer et~al.(2018)Kaizer, Alexander M and Hobbs, Brian P and
  Koopmeiners, Joseph S}]{kaizer2018multi}
Kaizer AM, Hobbs BP, Koopmeiners JS.
\newblock A multi-source adaptive platform design for testing sequential
  combinatorial therapeutic strategies.
\newblock Biometrics 2018;74(3):1082--1094.

\bibitem[{Hoeting et~al.(1999)Hoeting, Jennifer A and Madigan, David and
  Raftery, Adrian E and Volinsky, Chris T}]{hoeting1999bayesian}
Hoeting JA, Madigan D, Raftery AE, Volinsky CT.
\newblock Bayesian model averaging: a tutorial.
\newblock Statistical Science 1999;14(4):382--417.

\bibitem[{Fragoso et~al.(2018)Fragoso, Tiago M and Bertoli, Wesley and Louzada,
  Francisco}]{fragoso2018bayesian}
Fragoso TM, Bertoli W, Louzada F.
\newblock Bayesian model averaging: A systematic review and conceptual
  classification.
\newblock International Statistical Review 2018;86(1):1--28.

\bibitem[{Berry et~al.(2009)Berry, Donald and Wathen, J Kyle and Newell,
  Margaret}]{berry2009bayesian}
Berry D, Wathen JK, Newell M.
\newblock Bayesian model averaging in meta-analysis: vitamin E supplementation
  and mortality.
\newblock Clinical trials 2009;6(1):28--41.

\bibitem[{Wright(2009)Wright, Jonathan H}]{wright2009forecasting}
Wright JH.
\newblock Forecasting US inflation by Bayesian model averaging.
\newblock Journal of Forecasting 2009;28(2):131--144.

\bibitem[{Raftery et~al.(2005)Raftery, Adrian E and Gneiting, Tilmann and
  Balabdaoui, Fadoua and Polakowski, Michael}]{raftery2005using}
Raftery AE, Gneiting T, Balabdaoui F, Polakowski M.
\newblock Using Bayesian model averaging to calibrate forecast ensembles.
\newblock Monthly weather review 2005;133(5):1155--1174.

\bibitem[{Neuenschwander et~al.(2016)Neuenschwander, Beat and Wandel, Simon and
  Roychoudhury, Satrajit and Bailey, Stuart}]{neuenschwander2016robust}
Neuenschwander B, Wandel S, Roychoudhury S, Bailey S.
\newblock Robust exchangeability designs for early phase clinical trials with
  multiple strata.
\newblock Pharmaceutical statistics 2016;15(2):123--134.

\bibitem[{Hobbs and Landin(2018)Hobbs, Brian P and Landin,
  Rick}]{hobbs2018bayesian}
Hobbs BP, Landin R.
\newblock Bayesian basket trial design with exchangeability monitoring.
\newblock Statistics in Medicine 2018;37(25):3557--3572.

\bibitem[{Kaizer et~al.(2019)Kaizer, Alexander M and Koopmeiners, Joseph S and
  Kane, Michael J and Roychoudhury, Satrajit and Hong, David S and Hobbs, Brian
  P}]{kaizer2019basket}
Kaizer AM, Koopmeiners JS, Kane MJ, Roychoudhury S, Hong DS, Hobbs BP.
\newblock Basket designs: Statistical considerations for oncology trials.
\newblock JCO Precision Oncology 2019;3:1--9.

\bibitem[{Brown et~al.(2021)Brown, Roland and Fan, Yingling and Das, Kirti and
  Wolfson, Julian}]{brown2021iterated}
Brown R, Fan Y, Das K, Wolfson J.
\newblock Iterated multisource exchangeability models for individualized
  inference with an application to mobile sensor data.
\newblock Biometrics 2021;77(2):401--412.

\bibitem[{Kotalik et~al.(2021)Kotalik, Ales and Vock, David M and Donny, Eric C
  and Hatsukami, Dorothy K and Koopmeiners, Joseph S}]{kotalik2021dynamic}
Kotalik A, Vock DM, Donny EC, Hatsukami DK, Koopmeiners JS.
\newblock Dynamic borrowing in the presence of treatment effect heterogeneity.
\newblock Biostatistics 2021;22(4):789--804.

\bibitem[{Ling et~al.(2022)Ling, Sharon X and Hobbs, Brian P and Kaizer,
  Alexander M and Koopmeiners, Joseph S}]{ling2022calibrated}
Ling SX, Hobbs BP, Kaizer AM, Koopmeiners JS.
\newblock Calibrated dynamic borrowing using capping priors.
\newblock Journal of Biopharmaceutical Statistics 2022;p. 1--16.

\bibitem[{Ji and Wolfson(2022)Ji, Ziyu and Wolfson, Julian}]{ji2022flexible}
Ji Z, Wolfson J.
\newblock A flexible Bayesian framework for individualized inference via
  adaptive borrowing.
\newblock Biostatistics 2022;.

\bibitem[{Kaizer et~al.(2018)Kaizer, Alexander M and Koopmeiners, Joseph S and
  Hobbs, Brian P}]{kaizer2018bayesian}
Kaizer AM, Koopmeiners JS, Hobbs BP.
\newblock Bayesian hierarchical modeling based on multisource exchangeability.
\newblock Biostatistics 2018;19(2):169--184.

\bibitem[{Kotalik et~al.(2022)Kotalik, Ales and Vock, David M and Hobbs, Brian
  P and Koopmeiners, Joseph S}]{kotalik2022group}
Kotalik A, Vock DM, Hobbs BP, Koopmeiners JS.
\newblock A group-sequential randomized trial design utilizing supplemental
  trial data.
\newblock Statistics in medicine 2022;41(4):698--718.

\bibitem[{Morita et~al.(2008)Morita, Satoshi and Thall, Peter F and M{\"u}ller,
  Peter}]{morita2008determining}
Morita S, Thall PF, M{\"u}ller P.
\newblock Determining the effective sample size of a parametric prior.
\newblock Biometrics 2008;64(2):595--602.

\bibitem[{{R Core Team}(2021)}]{R}
{R Core Team}.
\newblock R: A Language and Environment for Statistical Computing.
\newblock R Foundation for Statistical Computing, Vienna, Austria; 2021,
  \urlprefix\url{https://www.R-project.org/}.

\end{thebibliography}

\end{document}